# A single radiation-hard pixel detector for proton dosimetry, imaging, energy resolution and beam monitoring

Nico Frédéric Brosda[1,*], Stéphane Higueret[2], Nicolas Arbor[2], Matilde Siviero[3], Maxime Hugues[3], Eric Frayssinet[3], Luisa Parschat[4], Christian Bäumer[4], Andreas Dirk Wieck[1], Jean-Yves Duboz[3]

1 Ruhr-University Bochum, Faculty of Physics and Astronomy, Experimentalphysik VI, D-44780 Bochum, Germany
2 Université de Strasbourg, IPHC, CNRS, UMR7178, F-67037 Strasbourg, France
3 Université Côte d'Azur, CNRS, CRHEA, F-06560 Valbonne, France
4 West German Proton Therapy Centre Essen, D-45147 Essen, Germany
*nico.brosda@rub.de

Keywords: Gallium Nitride, Proton therapy, FLASH, Semiconductor detectors, Radiation hardness, Dosimetry

## Abstract

Proton therapy enables highly conformal tumour irradiation by exploiting the Bragg peak. Clinical precision relies on detector systems that provide linearity, spatial resolution, radiation hardness, and, for ultra-high-dose-rate (FLASH) therapy, high dynamic range with sufficient temporal response. No compact all-in-one solution currently fulfils all these requirements.

Here, we present pixelated gallium nitride (GaN) arrays as a multifunctional platform. These detectors exhibit high linearity (proton currents 70–1700 pA $cm^{-2}$) and signal uniformity below 2% across 128 pixels. To probe GaN radiation hardness, we applied cumulative doses up to 6.7 MGy at energies from 25 MeV down to the Bragg-peak regime. Locally deposited dose dictates long-term signal loss, yet devices preserve linearity and spatial response.

We demonstrate 0.25 mm effective spatial resolution, validated against radiochromic film. Proton radiography resolves fine tissue structures, with the dose extrapolated to ~7 mGy per 1-ms frame in a prospective single-shot configuration. Energy-dependent responses match GATE/Geant4 simulations, and wedge measurements enable Bragg-peak tracking with an energy resolution below 10 keV in an optimised geometry. Real-time ms-scale beam monitoring enables live dose verification.

These results establish GaN pixel detectors as a scalable and radiation-hard platform, enabling dosimetry, beam monitoring and energy verification in one device.

# Introduction

Proton therapy delivers highly conformal doses to tumour tissue by exploiting the localised energy deposition of the Bragg peak[1,2]. The same property makes the delivered dose sensitive to uncertainties in particle range, arising from variations in patient anatomy, setup and beam delivery, and from the conversion of CT (Computed Tomography) images to relative stopping power[3]. These uncertainties are currently compensated by beam-specific safety margins and robust plan optimisation, which secure target coverage at the price of additional dose to healthy tissues[4]. Reducing this cost drives a shift from predicting the delivered dose towards measuring it: adaptive workflows tailor the plan to the daily anatomy[5,6], in-vivo verification assesses the delivered range[7] and particle-based imaging measures stopping power directly[8–10]. In parallel, ultra-high-dose-rate (FLASH) irradiation[11], proton arc therapy[12], spatial fractionation[13] and heavier or mixed ion species[14] extend delivery into regimes of intensity, timing, dose gradient and ionisation density where the linearity and stability of established instruments are no longer assured. Cost and availability remain additional challenges[15], motivating new compact accelerator concepts[16]. These developments share a reliance on measuring the beam itself, but differ by orders of magnitude in the temporal structure, instantaneous flux and spatial resolution at which that measurement must be performed.

Clinical delivery and further development of proton therapy rest on detector systems that characterise the beam during commissioning and quality assurance and monitor it during treatment[17]. Within the nozzle, segmented, radiologically thin detection planes provide real-time information on beam properties. Depth-dose profiles and lateral beam characteristics are commonly verified using ionisation chambers[18], CCD-coupled scintillators[19], silicon-based strip detectors[20,21] or amorphous silicon flat-panel arrays[22]. In particular FLASH pushes detectors beyond their dynamic range[23,24], motivating exploration of materials such as silicon carbide (SiC)[25–27], amorphous silicon[28], diamond[29,30] or hybrid-material scintillators[31,32]. Across these applications a recurring set of requirements emerges: high spatial resolution over a clinically relevant active area, real-time readout on the timescale of beam delivery, sufficient dynamic range with high linearity and sensitivity to the energy of the incident particles. No single device combines these within a compact, thin geometry that remains stable at therapeutic fluences. Related demands arise beyond the clinic at accelerators, in nuclear decommissioning and in space[33,34].

Semiconductor detectors offer a route towards this profile, combining intrinsic signal linearity, fast charge collection and compatibility with fine spatial segmentation[35]. However, conventional semiconductor materials face limitations: silicon detectors exhibit performance degradation at high particle fluences[35,36], while radiation-hard diamond has been demonstrated at small active areas[29,30,37], with large-area fabrication remaining a limiting factor. Further materials are under active investigation: gallium oxide shows promising radiation tolerance[38], perovskites achieve high sensitivity, though not yet combined with scalable large-area fabrication[39–41]. Among wide-bandgap semiconductors, SiC and gallium nitride (GaN) are distinguished by a mature, industrially established fabrication base, and SiC in particular has been characterised extensively for detector applications[42,43].

Gallium nitride shares a wide bandgap and high radiation tolerance[44–46] with SiC, while differing in a higher charge yield per unit thickness, which suits radiologically thin detectors. The heavy gallium constituent responsible for this yield also removes water equivalence, placing the requirement on the detector to have a calculable response. Although GaN has been investigated for ultraviolet[47], X-ray[48,49], gamma-ray[50], alpha-particle[51,52], and neutron[50,53] detection, its use in clinical proton instrumentation remains comparatively underexplored. Previous studies demonstrated GaN Schottky diodes for clinical-energy proton detection and range verification[54–56], and recent work identified GaN pin-diodes with a μm-scale active i-GaN layer as a favourable geometry for stability, signal-to-noise ratio and manufacturability[57]. Depending on bias conditions, GaN diodes can detect proton fluxes down to 5 protons per second[58] or provide a highly linear and stable response near zero bias, as used in this study.

Here we develop GaN-based detector structures from single diodes towards pixel arrays tailored to proton therapy. We characterise GaN pin-diode pixel detectors under proton irradiation from the Bragg-peak regime to clinical treatment energies, addressing the requirements identified above, from linearity and spatial resolution to energy sensitivity, temporal response and stability under accumulated dose. Monte Carlo calculations reproduce the measured response from fundamental energy deposition. We thereby establish GaN pin-diode arrays as a detector platform for beam monitoring, quality assurance and as a basis for extension towards range-sensitive measurements and proton imaging within a single device concept.

# Results and discussion

## Detector design and signal quality

Gallium nitride (GaN) pin-diode detectors were fabricated on GaN grown on 3-inch sapphire substrates and structured into pixel arrays with different geometries, enabling a systematic investigation of spatially resolved detector performance (Fig. 1a–g). Two principal designs were realised: linear diode arrays and two-dimensional matrices. The linear arrays consist of 128 individual diodes arranged with a pitch of 0.5 mm and lateral diode dimensions of either 0.25 mm or 0.5 mm orthogonal to the array direction (Fig. 1f,g). In addition, matrix detectors with pixel pitches of 0.5 mm and 1 mm were fabricated (Fig. 1b–d). Details of the fabrication process and diode structure are provided in the Methods section.

Charge signals generated in the diode structures were read out via direct wire bonding of individual pixels to a printed circuit board connected to an external readout circuit positioned outside the irradiation field (Fig. 1e). In the present study, a 128-channel low-noise current readout chip (AMS-OSRAM CT-common AS5900) was used, enabling simultaneous acquisition of signals from full detector arrays. Detector characterisation was performed using the continuous-wave proton beam of a research cyclotron with energies up to 25 MeV and beam currents of up to 15 nA[59–61], corresponding to intensities relevant for proton therapy applications. Additional details of the experimental configuration are described in the Methods section.

Response homogeneity across individual pixels was evaluated by measuring detector signals at overlapping beam positions, enabling a direct comparison of the signal current of neighbouring diodes. This approach both quantifies intrinsic response variations and yields correction factors for subsequent measurements. Figure 1h shows the signal distributions for linear arrays with diode widths of 0.5 mm (Fig. 1f) and 0.25 mm (Fig. 1g). In both cases, the histogrammed signals follow a Gaussian distribution with a response dispersion as low as 0.6% in the 0.5 mm diode array and below 2% overall. This excellent homogeneity observed across the array confirms the suitability of the fabrication process for scalable detector arrays.

The fabrication yield of the detector process is correspondingly high, with 255 out of 256 pixels operational in the investigated arrays (Fig. 1h, inset).

The linearity of the detector response was assessed by comparing the mean signal of the arrays over proton beam currents ranging from 70 to 1700 pA $cm^{-2}$ (1 pA $cm^{-2} \approx 6.24 \times 10^{6}$ protons $s^{-1}$ $cm^{-2}$) (Fig. 1i). A linear dependence on beam current is observed across the full investigated range, as quantified by the coefficient of determination ($R^2=0.99994$) for a proportional fit. Linearity versus proton current has previously been observed at the individual-diode level over several decades of intensity[57]. Here it is confirmed at the system level over more than an order of magnitude in beam current. The standard deviation of the signal scales with the square root of the mean response, consistent with Poisson statistics and indicating that the observed signal fluctuations are dominated by shot noise and generation-recombination noise.

This further indicates that the detector response is dominated by usual fundamental statistics, providing a robust basis for quantitative dosimetry.

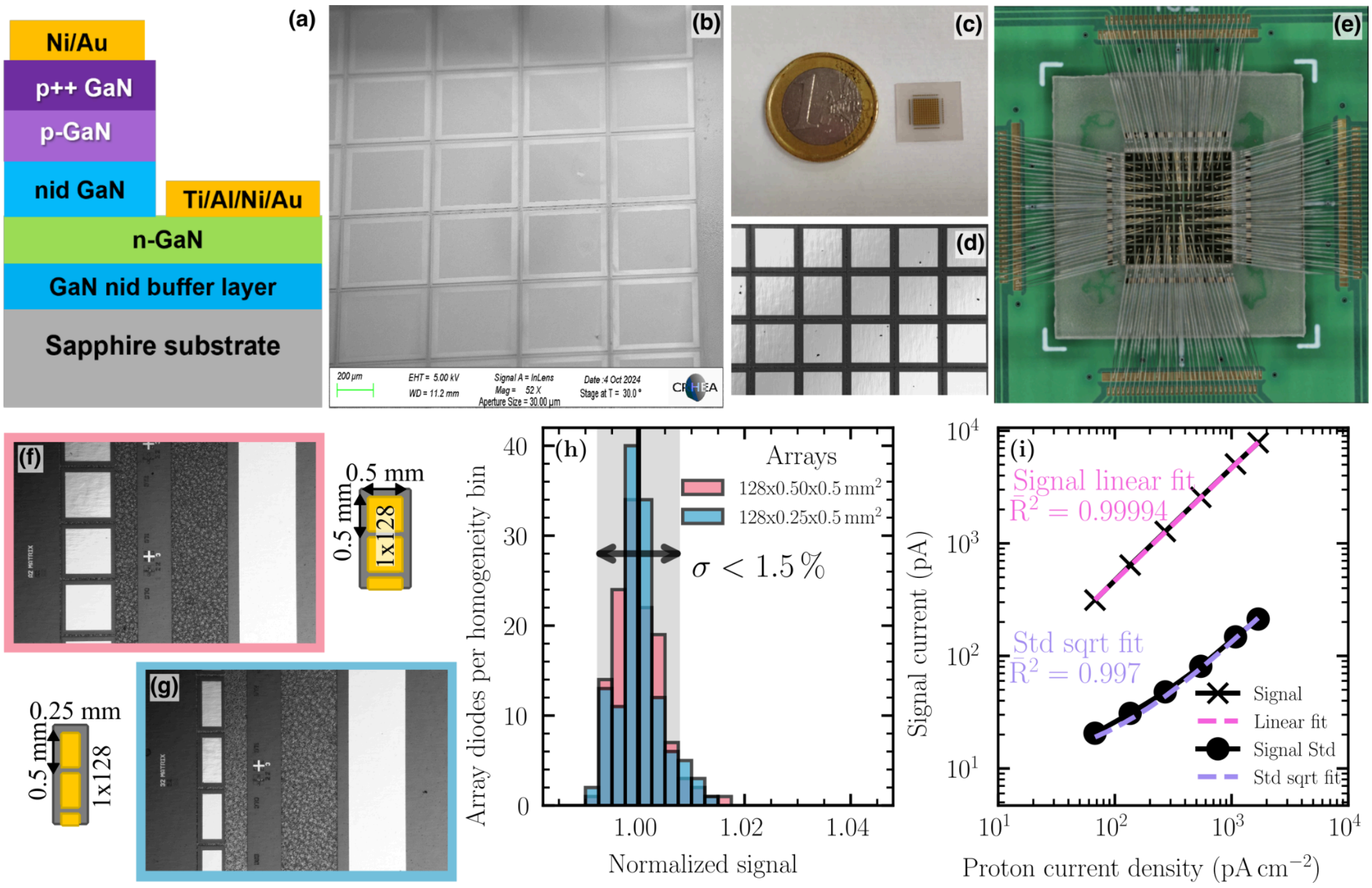

Fig. 1: Detector design and signal quality

**a,** Schematic of the pin-diode detector structure used. The layer thicknesses are specified in the Methods section. Dimensions are not to scale.

**b–d**, SEM images and photographs of GaN pin-diode detectors fabricated on sapphire substrates in different pixelated geometries, including two-dimensional matrix layouts with pixel pitches of 0.5 mm and 1 mm.

**e**, Photograph of the detector mounted on a printed circuit board with individual pixels wire-bonded to an external multichannel readout circuit positioned outside the irradiation field.

**f,g**, Linear GaN diode arrays composed of 128 pixels with a pitch of 0.5 mm and lateral diode dimensions of **0.5 mm** (**f**) and **0.25 mm** (**g**) orthogonal to the array direction.

**h**, Pixel-by-pixel response homogeneity measured using overlapping beam positions for the linear arrays shown in **f** and **g.** The distributions of normalised pixel responses follow Gaussian profiles with relative inhomogeneities below 2%. Inset: Homogeneity distribution over the array. The operational yield of the arrays proves to be near 100%, with 255 out of 256 pixels functional.

**i**, Linearity of the detector response as a function of proton beam current, shown for current densities ranging from 70 to 1700 pA cm$^{-2}$. The mean detector signal exhibits a proportional dependence on beam current (crosses and pink dashed line for the fit), while the mean of each channel's signal standard deviation, derived from 1000 recorded measurements per channel, follows square-root scaling, consistent with Poisson statistics (circles and fit as purple dashed line).

## Radiation hardness after multi-megagray exposure

Detectors operating in the vicinity of proton therapy beamlines are exposed to substantial cumulative radiation doses, ranging from milligray levels for potential imaging applications to tens of gray per patient treatment, with dose rates exceeding 40 Gy $s^{-1}$ in emerging FLASH therapy regimes[23,24]. To assess the radiation tolerance of the GaN detectors under extreme conditions, long-term irradiation experiments were performed. We first targeted a cumulative dose exceeding 2.4 MGy, corresponding to more than $5 \times 10^4$ patient treatments, at a constant dose rate of 43.6 Gy $s^{-1}$. The irradiation was carried out using 25 MeV protons, an energy below typical initial proton therapy energies (60–230 MeV) while remaining well above the proton energies encountered near the Bragg peak within the GaN detector volume.

The detector signal was continuously monitored throughout the 15 h long irradiation campaign , with the measured current and estimated dose shown as a function of irradiation time in Fig. 2a. Drops of the beam current on a minute timescale and a following readjustment are visible fluctuations in the signal trace. Over the course of the irradiation, the detector signal decreases by approximately a factor of three, from an initial value exceeding 8 nA to about 3 nA, while remaining well above zero. Notably, the signal degradation is non-linear in time but follows an exponential like evolution that will be analysed later on. In the final phase of irradiation, the signal appears to converge towards a non-zero response, with only marginal further degradation despite continued dose accumulation.

Despite the high cumulative dose of 2.4 MGy, the detector remains fully operational after irradiation. Following a cool-down period of approximately three months to allow for the decay of irradiation-induced activity, the detector response was re-evaluated. Figure 2b compares the signal linearity measured before irradiation and after the high-dose exposure. In both cases, the detector response follows the expected proportional dependence on beam current over a wide dynamic range. This preservation of linearity indicates that the fundamental charge-collection behaviour of the GaN diode remains intact even after severe irradiation.

To assess the spatial dosimetric functionality of the detector after irradiation, the response of diode arrays exposed within a 36 mm beam aperture was analysed. Due to the insertion of a 20 µm aluminium degrader foil in the beamline, the proton beam intensity exhibits a modest spatial inhomogeneity across the aperture, which is resolved by the detector array. Figure 2c shows normalised spatial response profiles measured before and after the irradiation campaign. While the exact beam shapes vary between measurements due to beam readjustments, both profiles display comparable signal ranges between the beam maximum and the aperture edges. This demonstrates that the detector retains its ability to spatially resolve dose distributions even after exposure to multi-megagray radiation levels.

Despite the pronounced signal reduction observed at extreme cumulative doses, the detector remains fully functional, preserving both linear response and spatial sensitivity. The apparent saturation of degradation at high dose suggests the formation of a stable defect population.

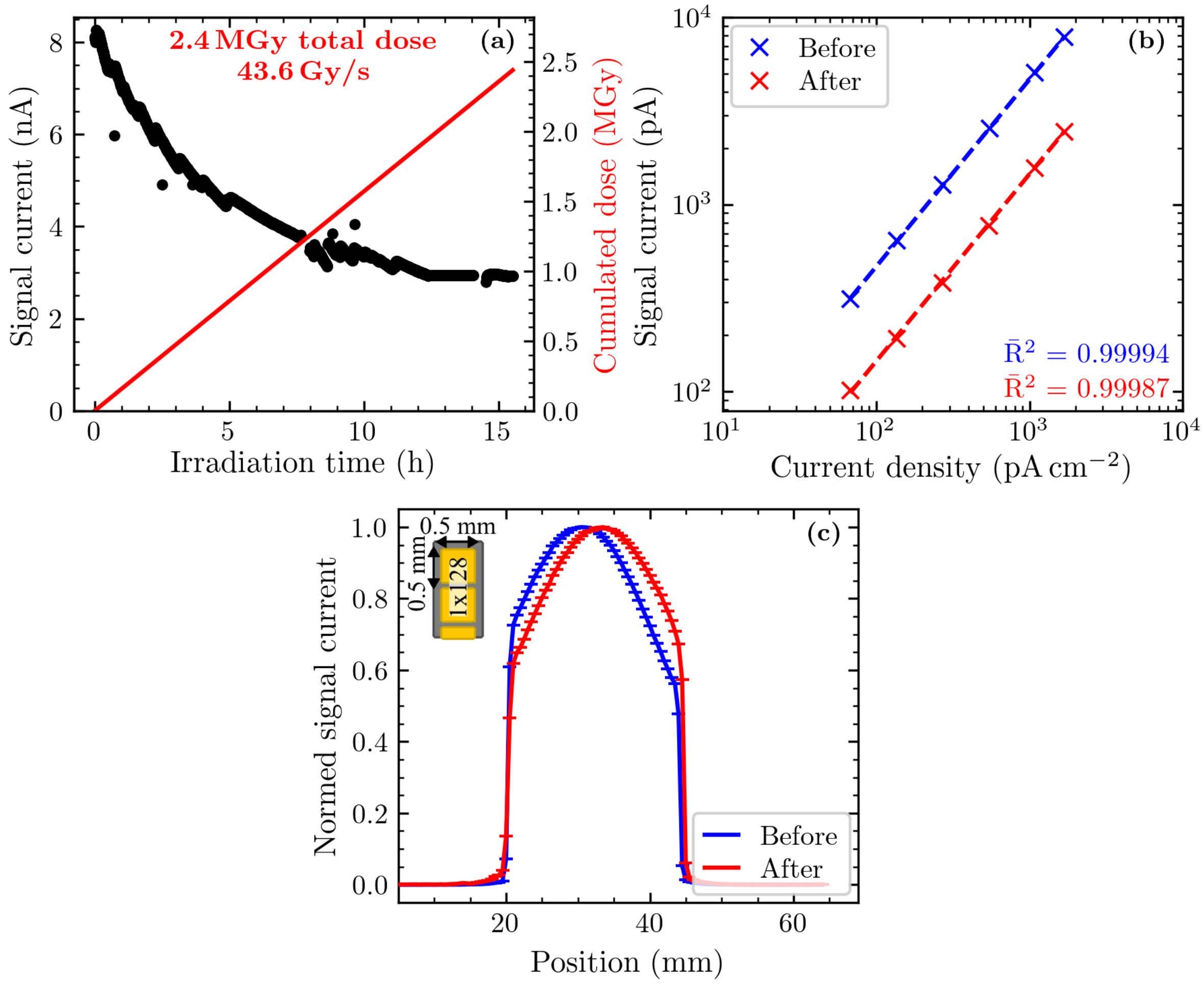

Fig. 2: Radiation hardness up to doses of 2.4 MGy

**a**, Detector signal recorded continuously during proton irradiation with 25 MeV protons as a function of irradiation time, with the accumulated dose indicated on the right vertical axis. The proton beam current was periodically verified and readjusted during the irradiation campaign, resulting in discrete steps in the signal trace. The cumulative dose reaches 2.4 MGy and was delivered at a dose rate of 43.6 Gy $s^{-1}$.

**b**, Detector response as a function of proton beam current measured before irradiation and after the irradiation campaign (following a cool-down period of approximately three months), showing the response linearity.

**c**, Normalised spatial response profiles of diode arrays measured before and after the irradiation campaign within a 36 mm beam aperture. While the spatial beam intensity distribution varies between measurements due to beam readjustments, the comparable signal range from the beam maximum towards the aperture edges demonstrates the preserved spatial dosimetric functionality of the detector after multi-megagray irradiation despite the signal loss of the irradiated diodes.

## Energy and dose dependence of degradation

A second long-term irradiation campaign was designed to probe radiation damage as a function of proton energy. While the incident beam energy was kept constant at 25 MeV, a wedge energy degrader inserted into the beamline modulated the energy reaching the detector across the range from 17 MeV down to below 1 MeV (Fig. 3a), thereby sampling conditions from well above the Bragg peak down to the stopping region. This configuration simultaneously allowed the cumulative dose to be extended up to 6.7 MGy at the highest-dose positions, probing detector degradation under even more extreme conditions than in the first campaign.

The signal recorded by the 2×64 diode array (0.5 mm pitch, 0.25 mm offset) displays a pronounced peak whose position does not directly correspond to the local proton energy (right y-axis, Fig. 3a). Rather, the measured signal reflects a convolution of the energy-dependent stopping power and the spatial proton density distribution shaped by the wedge degrader. To account for these coupled effects, the full experimental geometry was implemented in a Monte Carlo simulation, the GATE framework[62,63] based on Geant4[64], as described in the Methods (Simulation). All subsequent estimates of local dose, proton energy and proton fluence are derived from this simulation. The resulting simulated energy deposition profile reproduces the experimental signal well, as shown in Fig. 3a.

The irradiation campaign extended over more than 14 hours. Figure 3b shows the signal evolution of three representative diodes selected along the array, plotted against the cumulative proton fluence incident on each diode. The curves do not overlap: the diode exhibiting the largest signal reduction received fewer protons than the diode positioned towards the centre of the array, yet degraded more severely. Fluence alone therefore does not govern degradation. This indicates that the energy deposited per incident proton must be considered.

Replotting the same curves against the cumulative absorbed dose (Fig. 3c) brings them substantially closer together and reveals the scale of total detector exposure at MGy doses. Absorbed dose counts all deposited energy, the overwhelming majority of which goes into ionisation and does not displace lattice atoms. The quantity expected to govern displacement damage is instead the non-ionising energy loss (NIEL), the fraction of the stopping power transferred to nuclear rather than electronic collisions, and the corresponding displacement damage dose $D_d$[35,65].NIEL values for GaN were computed[66] and folded over the simulated proton spectrum.

To quantify the improved overlap while mitigating the influence of temporal beam fluence adjustments, the signal curves are fitted with a Kohlrausch stretched-exponential model as indicated in Fig. 3d:

$$S(d) = 1 - A \cdot (1 - exp\left(-\left(\frac{d}{k}\right)^{\beta}\right). \quad (1)$$

Here $A$ denotes the asymptotic fractional signal loss, β is the stretching exponent governing the shape of the decay, and $k$ is a damage scaling factor. Parameters $A$ and β are matched among all curves,with $A = 0.82$ and $\beta = 0.74$. The $k$ factor is allowed to vary along the array, and thus with proton energy (Fig. 3d, Inset).
A constant $k$ factor would show that the NIEL analysis is correct and sufficient to explain the degradation. Here, the $k$ factor varies with energy, increasing from 2 to 12

MeV, and then decreasing. This signals that the description of NIEL alone is not capturing all effects contributing to the degradation, in particular at the lowest energies where the $k$ variation is largest. This energy dependence exceeding the NIEL scaling can have several contributions not clearly identifiable without further measurements. A possible reason lies in the microscopic nature of the energy-deposition processes that differ fundamentally near the Bragg peak: slower protons may produce more clustered defect configurations than fast ones. In addition, the proton range is rendered differently across the diode array with changing energy, producing a depth-dependent damage profile that might impact different parts of the diode structure. Considering the proton fluence we do not expect retained hydrogen affecting Mg acceptor activation in p-GaN to be relevant. Deconvolving these contributions requires dedicated experiments and is left for future work.

The linearity of the diode array was evaluated before and after irradiation across all array positions, as shown in Fig. 3e. The mean response across all array positions remains linear throughout. The normalized root-mean-square errors of the linearity fits remain below 2% for all channels, both before and after irradiation. No systematic degradation of linearity can be observed in the diodes that received the highest cumulative doses. The preservation of linear response across all damage levels implies that signal loss due to irradiation can be corrected by a simple calibration measurement performed before and after high-dose exposure, as demonstrated in Fig. 3f. After correcting for the per-channel degradation, the image recovers the contrast that was lost near Bragg-peak proton energies (Fig. 3f).

Long-term degradation is well captured by NIEL, and to within tens of percent by total dose. For clinical use the signal loss under routine exposure is decisive. Considering the steepest part of the device degradation at low doses, the loss rate is in linear approximation at maximum 40 % over 1 MGy total dose. This equals a loss rate of 0.04 %/kGy, far lower than reported values for silicon diodes under proton exposure[35,36]. For clinical use the response remains effectively stable, enabling a usage also under higher dose loads.

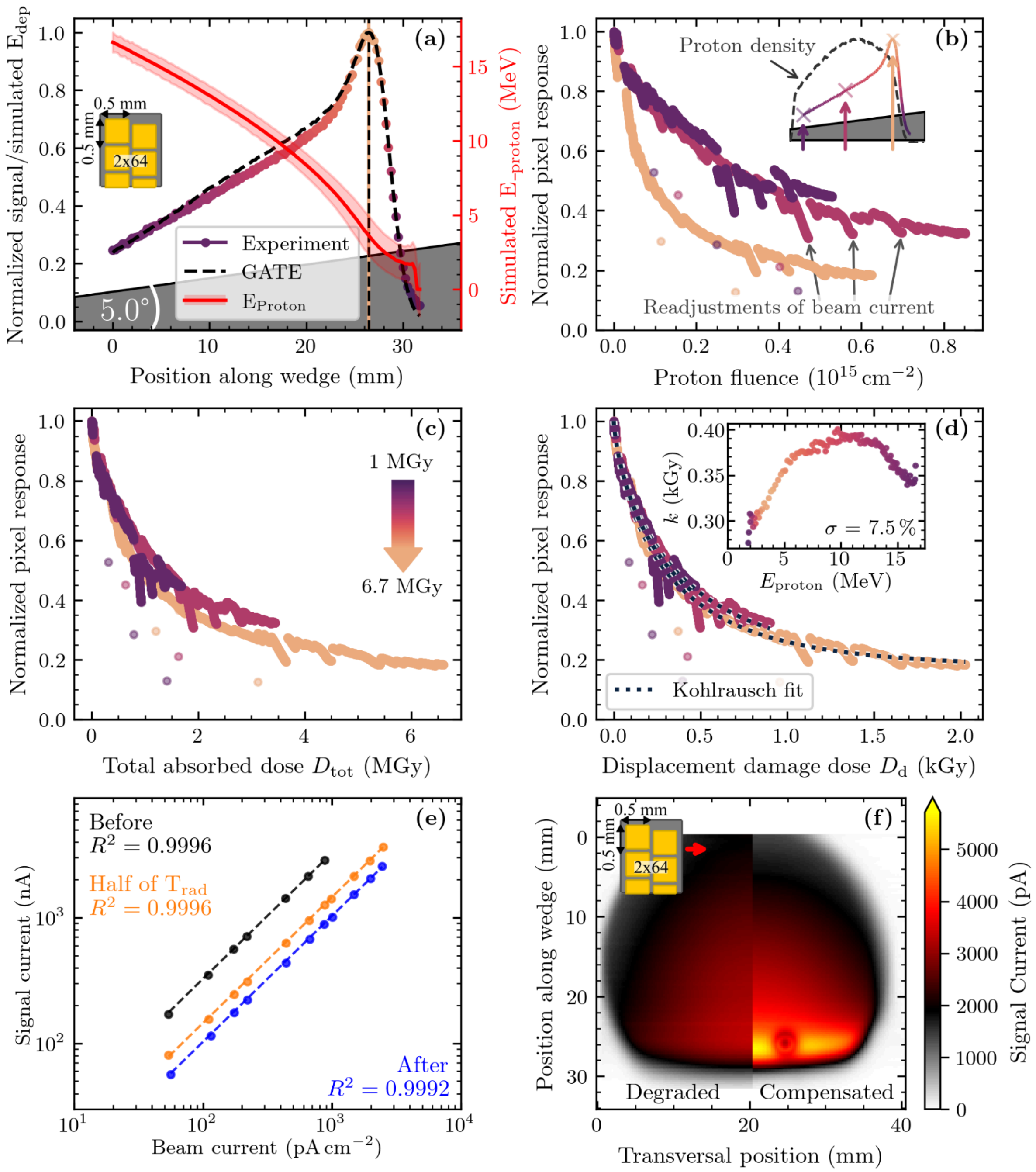

Fig. 3: Energy-dependent irradiation damage

**a**, Detector signal as a function of position along the diode array, recorded with the wedge energy degrader inserted into the beamline (left y-axis). The GATE-simulated energy deposition profile (black dashed line) reproduces the measured signal well. The right y-axis (red) shows the corresponding proton energy arriving at each detector position. The shaded band around the simulated energy curve represents the total energy spread at each detector position, arising from the convolution of the intrinsic beam energy spread, straggling within the wedge degrader and the finite pixel size.

**b,** Signal evolution of three representative diodes as a function of cumulative proton fluence incident on each diode, as estimated from the simulated proton density distribution. The curves do not superimpose, motivating the exposure scaling of **c** and **d**. Short-timescale drops in signal reflect beam current losses that were regularly checked and corrected during the irradiation campaign. Inset: simulated proton density distribution across the beam aperture; markers indicate the positions of the three selected channels.
**c,** The same three channels replotted against cumulative absorbed dose per diode. Agreement improves but a systematic energy dependence remains.
**d,** The same three channels replotted against displacement damage dose $D_d$ (NIEL × proton fluence). Dashed lines show stretched-exponential fits according to Eq. 1. Inset: damage constant k of the same fits against proton energy, for the 121 channels exceeding 1 MGy absorbed dose.
**e,** Linearity before, after half the radiation time $T_{rad}$ and after the irradiation campaign, demonstrating preserved proportional response across the tested dynamic range.
**f,** Image of the aperture with the wedge taken after the irradiation campaign. The left half shows the resulting array response as is, while the right half was corrected for the signal decrease of each channel. The hole in the lower right corner was drilled into the wedge for alignment control.

## Spatial resolution and radiochromic film validation

The spatial resolution of the GaN detector arrays was investigated using linear arrays with different pixel dimensions, corresponding to the detector geometries shown in Fig. 1f,g. To extend the effective sensitive area, the detector was mounted on a translation stage and scanned across the beam in discrete steps. A 3D-printed logo phantom was placed upstream of the detector (Fig. 4e) to generate spatially structured proton intensity patterns. The phantom contains curved and diagonal features, providing a stringent benchmark for spatial resolution and enabling the identification of potential signal distortions or reconstruction artefacts.

Figure 4a shows the spatially resolved detector response obtained with a linear array consisting of 128 diodes with lateral dimensions of 0.5 mm × 0.5 mm. The shape of the logo is clearly resolved, demonstrating the capability of the GaN detector to reproduce spatial dose variations. Reducing the diode dimension to 0.25 mm in the direction perpendicular to the array and adapting the linear translation step size (Fig. 4b) results in a visibly improved spatial resolution, particularly for curved and diagonal features of the logo. This improvement is accompanied by a corresponding reduction in signal amplitude, as halving the diode area leads to approximately half the collected charge per pixel. The fabrication process readily supports such smaller diode dimensions, enabling a flexible trade-off between spatial resolution and signal strength.

Beyond reducing pixel size, spatial resolution can be further enhanced by oversampling the beam using detector geometry. Figure 4c illustrates a configuration combining two 64-pixel arrays with a pitch of 0.5 mm and a half-pitch offset. This simple form of super-resolution improves the apparent spatial resolution without shrinking the pixels. Unlike the previous case, resolution here improves along the array direction without sacrificing signal amplitude, at the expense of a halved imaging field.

To validate the spatial resolution obtained with the GaN detector arrays, measurements were compared with radiographic projections acquired using radiochromic film, which offers a spatial resolution better than 25 μm and is effectively free of pixelation effects. The radiochromic film was placed directly in front of the phantom under otherwise identical irradiation conditions, at a reduced beam current. As the comparison focuses only on spatial contrast and not on absolute dose, the film was not dose-calibrated, and only normalised signal distributions were considered.

Figure 4 compares the spatial response obtained with the GaN detector arrays (4a–c) to the corresponding radiochromic measurement (4d), showing good qualitative agreement in the reproduction of the logo features. Fine structures visible in the GaN measurements are also present in the radiochromic reference, confirming that they originate from the phantom rather than from detector artefacts. Quantitative comparison of selected line profiles (Fig. 4f) demonstrates that the spatial extent and contrast of the features measured with the GaN detector closely match the film reference. These results confirm that the GaN detector arrays deliver a spatially accurate response.

The agreement between GaN detector measurements and radiochromic film images demonstrates that the achievable spatial resolution is primarily governed by detector geometry and sampling strategy rather than intrinsic material limitations. This scalability enables straightforward optimisation of spatial resolution through pixel size reduction or oversampling, depending on the targeted application.

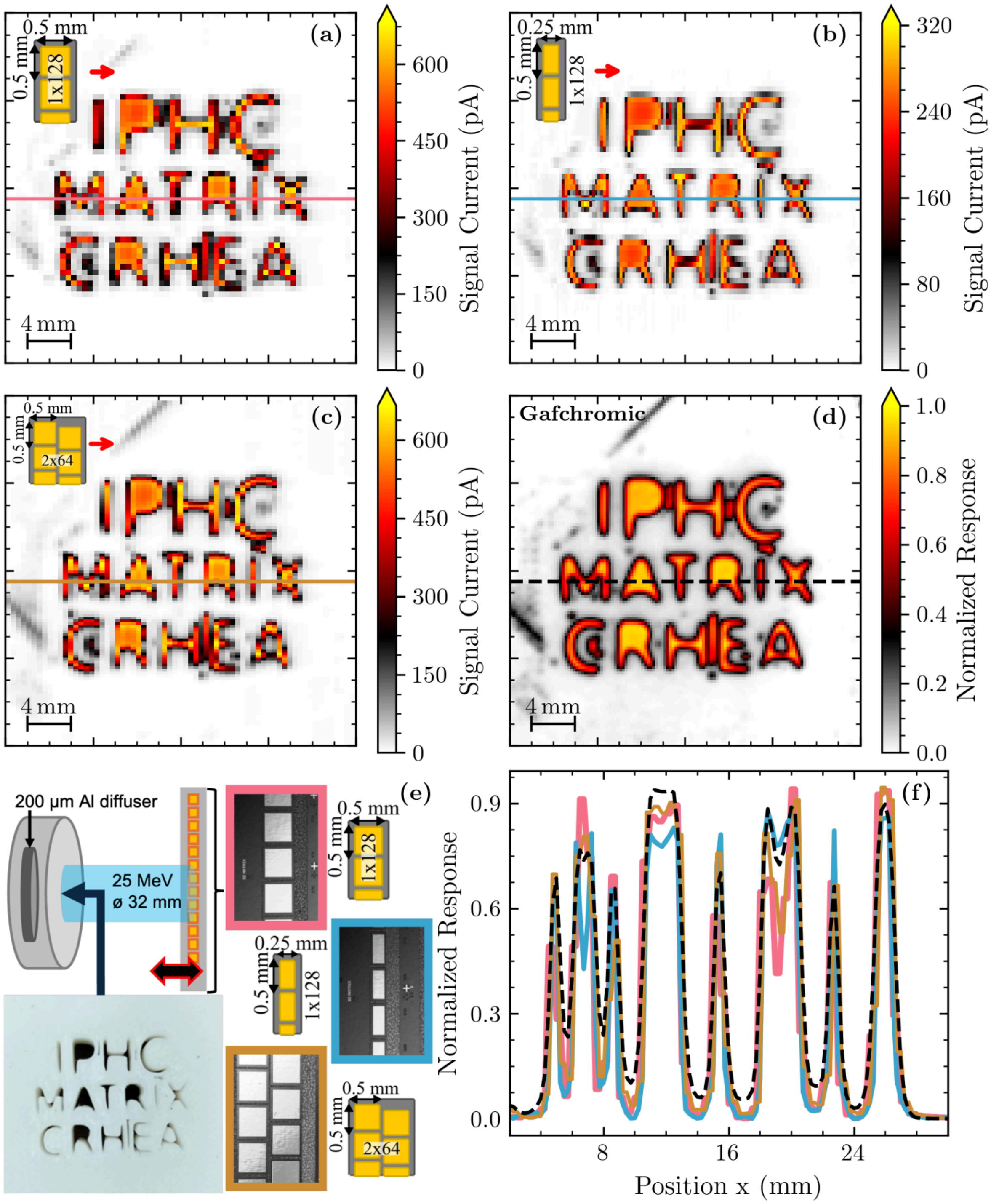

Fig. 4: Spatial resolution and radiochromic film validation

**a**, Spatial response measured with a linear GaN diode array consisting of 128 pixels with lateral dimensions of 0.5 mm × 0.5 mm. The red arrow shows the translation direction. The indicated line marks the position of the extracted profile used for comparison in Fig. 4f.

**b**, Corresponding spatial response measured with a linear GaN diode array consisting of 128 pixels with reduced lateral dimensions of 0.25 mm × 0.5 mm. Finer spatial features, particularly curved and diagonal letter segments, are more sharply resolved, demonstrating improved spatial resolution at the cost of reduced signal amplitude per pixel.

**c**, Corresponding spatial response measured with a linear GaN detector consisting of two 64-pixel arrays with lateral dimensions of 0.5 mm × 0.5 mm and sub-pitch translation steps, illustrating an oversampling-based super-resolution approach that enhances spatial details without modifying the detector geometry.
**d**, Radiographic projection of the same phantom acquired using radiochromic film placed directly in front of the phantom. The indicated line marks the profile position used for spatial resolution comparison in Fig. 4f. The film was not dose-calibrated, and only normalised signal contrast is considered.
**e**, Schematic of the measurement setup, showing the phantom inserted in the beamline, with microscopy images of the three array geometries used.
**f**, Comparison of normalised line profiles extracted from **a–d** (line colours correspond to those in Fig. 4a–d), demonstrating good agreement in the spatial extent and contrast of the features. The array with smaller pixel dimensions resolves finer spatial details, while both GaN detector measurements closely reproduce the reference profile obtained with radiochromic film.

## Application to pre-clinical radiography

As a proof-of-concept for spatially and energetically resolved measurements in a biomedical context, proton radiography of biological tissue was performed using a mouse foot (Fig. 5a). The complex structure of the specimen, including bone and surrounding soft tissue, requires both sufficient spatial resolution and sensitivity to energy-dependent proton attenuation. Measurements were carried out using an 11 × 11 GaN diode 2D array with a pixel pitch of 0.5 mm. To cover the full extent of the sample, the detector was translated in both lateral directions, and the resulting data were combined to form a two-dimensional radiographic projection.

Figure 5b shows the resulting proton radiograph of the mouse foot. Bone elements and variations between different tissue regions are clearly resolved. The displayed signal range was restricted to enhance contrast, and the colour scale was chosen to qualitatively match the convention used in computed tomography. For the applied irradiation conditions, 25 MeV protons at a beam current density of around 345 pA $cm^{-2}$, the estimated dose delivered by a single 1 ms measurement frame corresponds to approximately 7 mGy in a prospective single-shot detector configuration. The experiment was acquired by scanning the detector; the dose estimate therefore indicates the expected frame dose for a full-field detector with equivalent sensitivity.

These results demonstrate that GaN pixel detectors can resolve biologically relevant tissue contrast using low-energy protons at dose levels compatible with radiographic applications. The prospect of obtaining such contrast in a single, short exposure highlights the potential of GaN-based detectors for proton imaging scenarios where minimising delivered dose is essential.

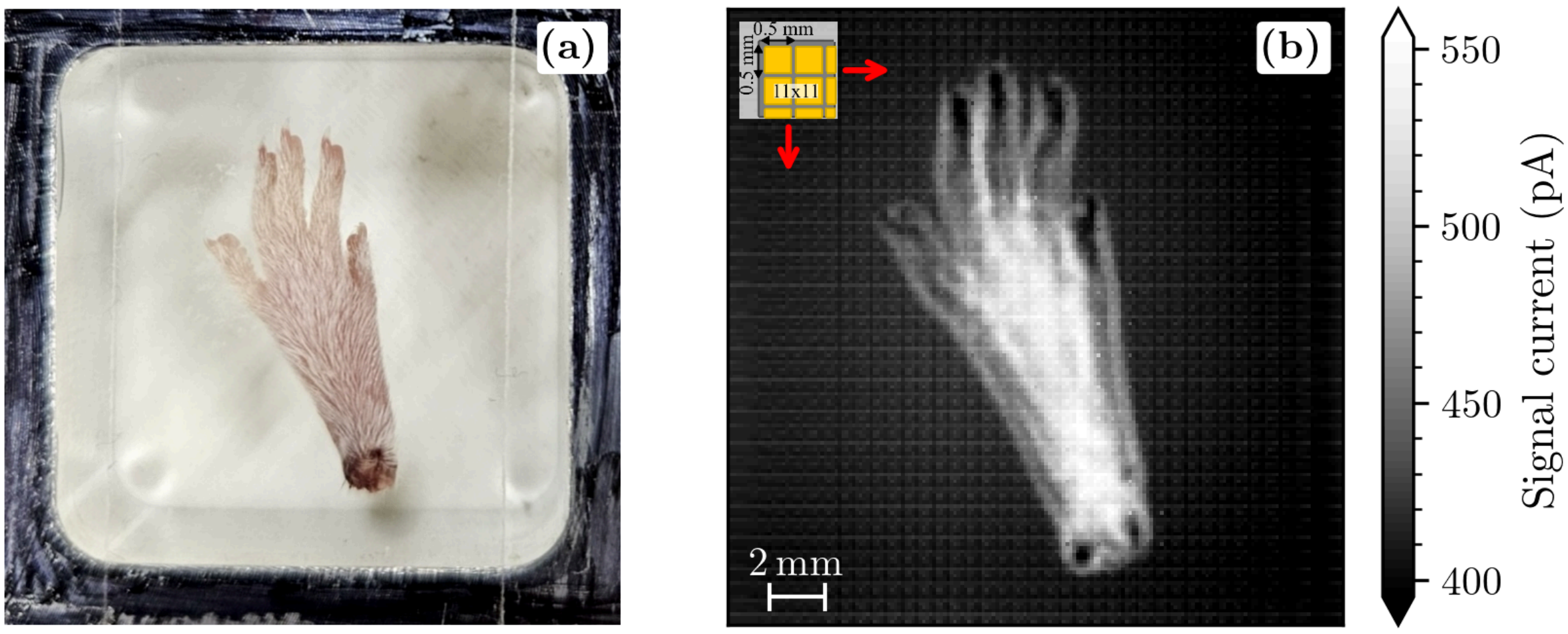

Fig. 5: Proton radiography of biological tissue
**a**, Photograph of the mouse foot specimen used for the proof-of-concept radiographic measurement, illustrating the complex biological structure requiring both spatial and energy-sensitive detection.
**b**, Proton radiograph of the mouse foot acquired with an 11 × 11 GaN diode array with a pixel pitch of 0.5 mm, translated in both lateral directions to cover the full sample area. Bone structures and soft-tissue contrast are clearly resolved. The signal colour scale is chosen to qualitatively match standard computed tomography conventions. The estimated dose delivered by a single 1 ms exposure is approximately 7 mGy.

## Energy-dependent response and Bragg-peak tracking

As demonstrated in the radiographic measurements of biological tissue, the GaN detector response exhibits contrast arising from variations in proton energy. To assess this energy dependence quantitatively, the detector response was measured as a function of proton energy and compared with simulated energy deposition. The corresponding energy deposition within the detector was simulated using the Monte-Carlo software GATE[62,63]. To enable direct comparison between measurement and simulation, the measured detector signal was normalised to the response per incident proton, accounting for the local current density (Fig. 6a). The detector response decreases as the proton energy increases, which is expected since higher-energy protons exhibit a lower stopping power and therefore deposit less energy per unit length in the sensitive volume.

Figure 6b compares the normalised detector signal to the simulated energy deposited in the active 1 µm GaN layer for different proton energies. Within the uncertainty of the energy modulation, the detector response is proportional to the deposited energy. The response remains monotonic over the full investigated energy range, preserving energy-dependent contrast. Absolute response values can also be interpreted using the Klein rule, in which the number of electron-hole pairs generated in a semiconductor is approximated by the deposited energy divided by three times the bandgap energy[67]. Under this estimate, 50 keV of deposited energy produces approximately 5000 electron-hole pairs. The response in Fig. 6b corresponds to approximately 4000 collected charge carriers, indicating a collection efficiency of about 80% in the pin-diode. This non-unity collection efficiency is consistent with a depletion region smaller than the nominal undoped layer under proton irradiation, as discussed previously[58].

Measurements at the West German Proton Therapy Centre Essen (WPE) extended the energy range to therapeutic proton energies from 100 to 226.7 MeV behind 2 cm of water-equivalent material (inset Fig. 6b). The data are consistent within an order of magnitude with the trend predicted from measurements at lower energies. As the measurements were recorded at different facilities under different beam conditions, a more quantitative comparison should be interpreted with caution. Nevertheless, the agreement is a promising indication of the detector's suitability for clinical proton energies.

The capability of the detector to resolve proton energy differences was further explored using a wedge-shaped absorber with a 14° opening angle placed upstream of the detector. This geometry introduces a spatially varying energy loss, enabling the Bragg peak of the proton beam to be resolved across the detector array (Fig. 6c). The measured signal peak reflects a convolution of proton stopping power and local proton density, as already explained. Changing the incident proton energy systematically shifts the signal peak position along the detector array, enabling range verification[68]. For the present wedge geometry and pixel pitch, the energy resolution estimated from the shift required to move the maximum signal peak from one diode to the next is approximately 350 keV.

This resolution can be further improved by optimising the wedge geometry and reducing the effective pixel pitch. A more refined analysis of the Bragg-peak position was performed in a second measurement (Fig. 6d) using an optimised PEEK wedge with a smaller 5° opening angle and an effective pixel pitch of 0.25 mm. The Bragg-peak position was determined either by fitting the peak or by taking its derivative and interpolating the zero crossing; both approaches yielded comparable results. Repeated spectra at the same energy gave a peak-position uncertainty of about 0.1 pitches. Combining this uncertainty with the measured peak-position dependence on proton energy (15.8 pitches per MeV) gives an energy resolution of approximately 7 keV. We therefore conservatively claim an energy resolution below 10 keV for the optimised setup.

The GATE simulations reproduce the measured detector response profiles in both wedge configurations well, accurately predicting the signal peak positions across the array for all investigated incident proton energies (Fig. 6c, d). The close correspondence between simulated and measured peak positions demonstrates that the underlying signal generation process in the GaN detector is well understood and can be reliably modelled. The simulation also captures the characteristic shape of the signal peak behind the wedge, including the energy spread visible in the peak profile, indicating that straggling and scattering processes are sufficiently modelled. The simulation framework can be used to predict detector responses and optimise measurement geometries for different wedge configurations, pixel pitches and proton energy ranges.

The linear relationship between detector signal and deposited energy over the investigated proton-energy range supports the use of GaN detectors for quantitative, energy-sensitive measurements. The extension of this response to therapeutic proton energies, consistent within an order of magnitude with the trend established at lower energies, shows the potential of GaN detectors within the clinical environment. In addition, the wedge measurements demonstrate that spatially resolved GaN detector arrays can reliably detect Bragg-peak shifts and thus resolve proton-energy differences. In the optimised configuration, this approach enables an energy resolution below 10 keV, highlighting the potential of GaN detectors as compact tools for energy-resolved beam diagnostics and range verification.

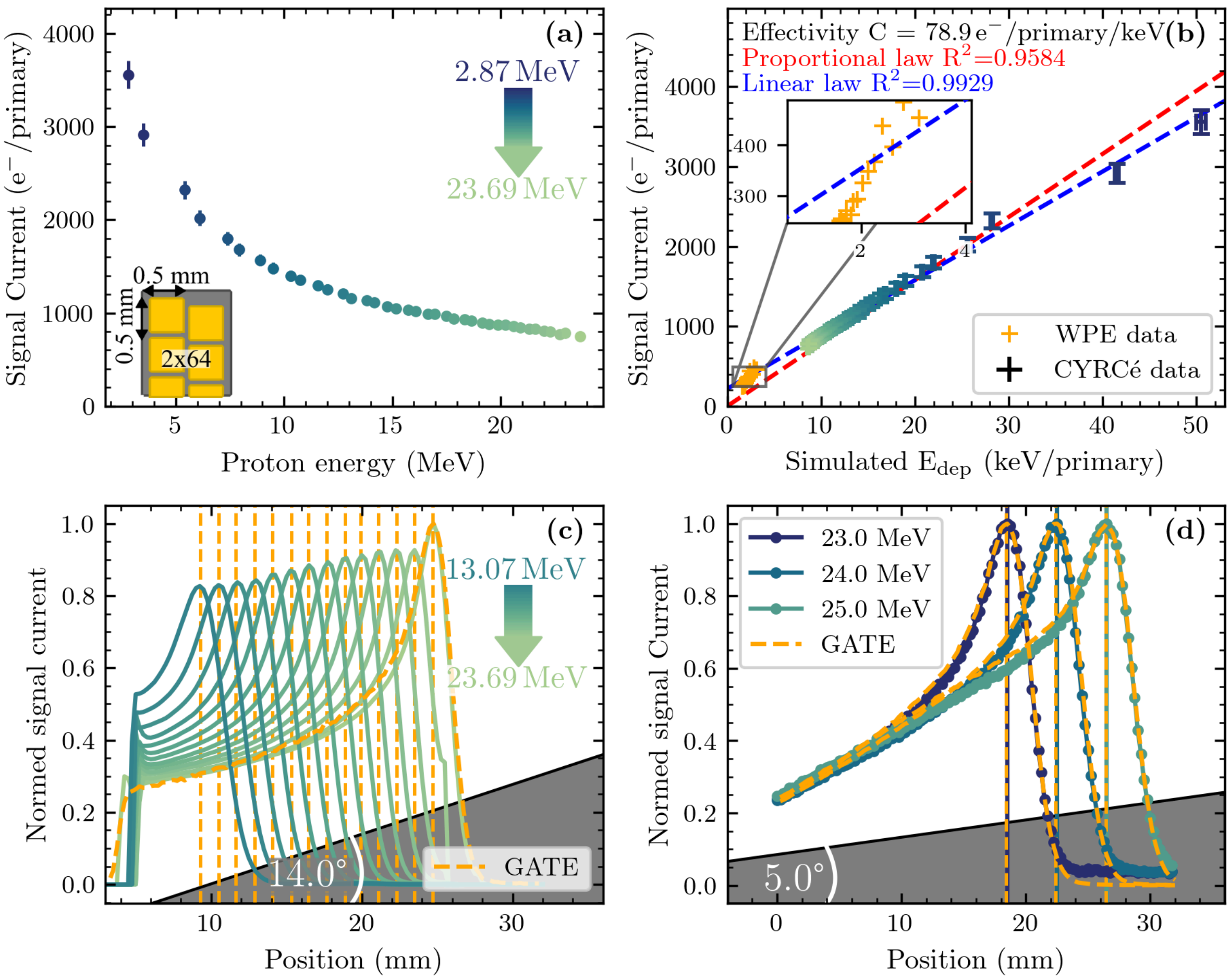

Fig. 6: Energy-dependent response and Bragg-peak tracking

**a,** Detector signal current as a function of incident proton energy, varied by inserting aluminium absorber foils of different thicknesses into the beam. The signal was normalised to the incident proton current density. Measurements were performed with the same detector over an energy range of 2.87–23.89 MeV.

**b,** Normalised signal current as a function of simulated energy deposition in the active GaN detector layer. Linearity of the detector response is assessed by fitting a proportional dependence, from which the efficiency of converting deposited energy into measured signal current is extracted. Inset: measurements performed at a proton beam therapy facility, extending the energy range from 100 to 226.7 MeV behind 2 cm of water-equivalent material.

**c,** Spatially resolved detector response measured with a polypropylene wedge absorber (14° opening angle). The recorded signal reflects a convolution of the Bragg-peak stopping-power profile and the local proton density across the detector array. Varying the incident proton energy via aluminium absorber foils of different thicknesses produces a systematic shift of the Bragg-peak position along the array. Measured profiles (individual points connected by straight lines) are compared with the peak positions obtained by GATE Monte Carlo simulations (dashed, orange line; curve overlay only shown for the highest energy). The remaining ~1 mm mismatch between simulation and experiment is consistent with the positioning accuracy of the wedge and detector; all simulated peak positions were shifted by the same constant offset.
**d,** Spatially resolved detector response measured with a shallower PEEK wedge absorber (5° opening angle), providing finer energy dispersion across the array. Measured profiles are compared with GATE Monte Carlo simulations of the experimental geometry. The incident energies were varied directly at the cyclotron.

## Real-time beam monitoring

To assess the suitability of GaN pixel detectors for real-time beam monitoring during proton therapy, we investigated their temporal response under continuous in-beam operation. A fixed two-dimensional detector array with a pitch of 1 mm, covering an active area of 11 × 11 $mm^2$, was statically placed in the unhomogenised proton beam. This configuration reflects a scenario in which the detector remains permanently inserted into the beamline to monitor dose delivery, beam stability during treatment, or beam-shape tuning.

The unhomogenised beam spot was monitored continuously using a detector integration time of 1 ms. Figure 7a illustrates representative observations on a seconds timescale. During the measurement, a profiling wire integrated into the beamline was moved to acquire beam profiles. The motion of this wire is clearly resolved in the detector signal, demonstrating the capability of the GaN array to track dynamic changes in beam position and shape in real time. For clarity, the static background contribution of the beam spot has been subtracted; the full sequences are shown in Supplementary Videos S1 and S2, without and with this subtraction, respectively.

At the millisecond timescale, temporal fluctuations of the beam hotspot were observed (Fig. 7b); the full sequence is shown in Supplementary Video S3. These fluctuations exhibit a characteristic frequency of approximately 50 Hz and had not been resolved by standard diagnostics used before. This observation highlights the sensitivity of the detector to fast beam dynamics and intrinsic temporal instabilities. In the present setup, the maximum achievable frame rate is limited by the readout electronics; for the presented results we used an integration time of 1 ms, while the integration time can be reduced to 50 µs with the same readout electronics.

The demonstrated millisecond-scale temporal resolution and sensitivity to fast beam fluctuations highlight the suitability of GaN pixel detectors for real-time beam monitoring during treatment. The present temporal performance is limited by readout electronics rather than the sensor material, so faster acquisition should be feasible. This suggests GaN detectors may be applicable to FLASH beams: for example, at dose rates 20× higher, the integration time could be shortened from 1 ms to 50 µs while preserving the integrated signal demonstrated here, giving a temporal resolution sufficient to resolve FLASH pulses.

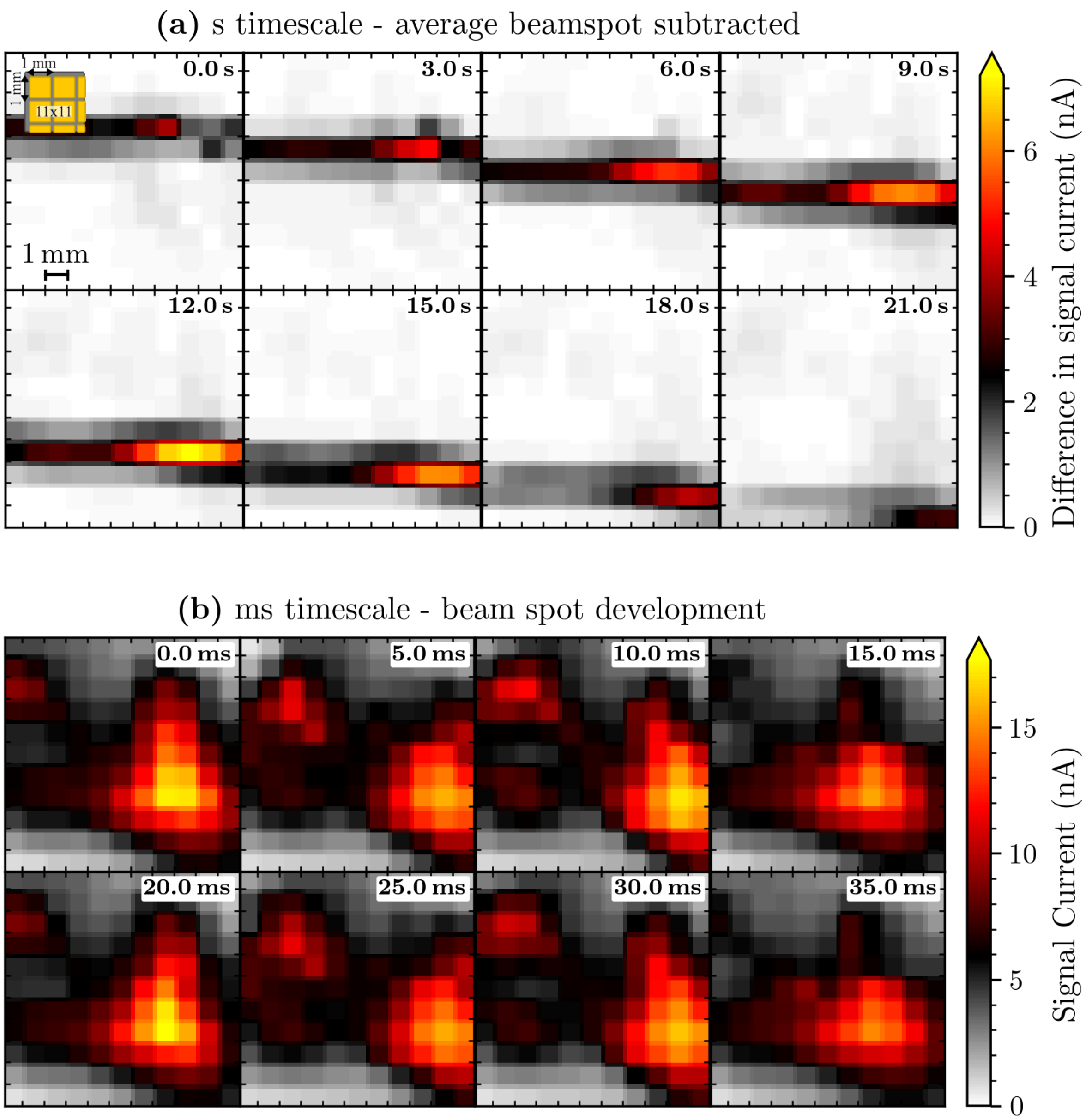

Fig. 7: Real-time beam monitoring

Frames of a movie taken by a fixed 11 × 11 $mm^2$ detector array. The observations on different timescales demonstrate the detector's capability as a live monitoring device. The full sequences are provided as Supplementary Videos S1–S3.

**a,** Movement of a wire inside the beamline used for beam profiling (Supplementary Video S2; the same sequence without background subtraction is Supplementary Video S1). The average beam spot shown in **b** is subtracted from the signal, and the absolute value of the difference is displayed. Behind the wire the signal arriving at the detector is lowered, with higher signal amplitudes where the beam is more intense. Each displayed frame is averaged over 0.5 s of measurement.

**b,** Development of the unhomogenised proton beam spot on a millisecond timescale (Supplementary Video S3). The unhomogenised beam shows different hotspots, depending on the collimation settings. A periodic fluctuation of 20 ms (top and bottom images, separated by 20 ms, are similar) can be observed within the beam spot, corresponding to a frequency of 50 Hz. Each frame corresponds to a 1 ms integration cycle.

## Conclusion

Taken together, the results presented here demonstrate that GaN detector arrays are a versatile platform for proton beam diagnostics under clinically relevant conditions. The combination of linear and homogeneous signal response, sub-millimetre spatial resolution, millisecond temporal resolution and sustained functionality after multi-megagray irradiation addresses key requirements for beam monitoring, quality assurance and proton radiography within a single detector concept. Beyond static dosimetry, the demonstrated capability for in-beam, real-time operation and sensitivity to fast beam dynamics highlights the potential of GaN detectors to provide information that is not accessible with conventional integrated or film-based systems. By fully or partially removing the substrate, quasi-transparent detector arrays could be fabricated, allowing for real-time beam monitoring during treatments.

The scalability of GaN device fabrication and its compatibility with pixelated layouts provide a pathway to larger-area, application-specific systems, motivating further research towards clinical translation. In this context, GaN complements ongoing developments in other wide-bandgap semiconductors, such as SiC, by expanding the material landscape for radiation-hard detector technologies. The present work establishes GaN as a platform for multifunctional detector systems in proton therapy, with potential impact ranging from routine beam quality assurance through advanced imaging and future high-dose-rate regimes to applications beyond the clinic.

# Methods

### *Detector fabrication*

The wafers are grown on 3-inch sapphire substrates (thickness 750 µm) using metal-organic chemical vapour deposition (MOCVD) in an Aixtron 6 Closed Coupled Showerhead. After a 2 µm GaN buffer layer, an n-doped GaN layer (2 µm) is grown followed by a nominally undoped (nid) GaN layer of 1–2 µm, depending on the sample. To form pin diodes, $p^+$ (0.5 µm) and $p^{++}$ GaN (30 nm) layers are grown on top.

The arrays of diodes are fabricated in the clean room. UV photolithography is used. The mesa is etched down to the n layer using a $BCl_3/Cl_2$ inductively coupled plasma (ICP-RIE). Then an n-contact (Ti/Al/Ni/Au) is deposited on the bottom of the mesa (n-GaN) and annealed at 750 °C for 30 s. The p-contact of (Ni/Au) is deposited on the top of the mesa and annealed at 500 °C for 4 min. At the end of the fabrication process the wafer is cut using a diamond blade to separate each array.

The active i-GaN layer thickness differed between wafers used for fabrication: the three arrays used for the homogeneity, spatial-response and energy-dependent measurements had a 1 µm active layer (for internal reference: fabricated from CC2232A), the one used for the linearity and radiation-hardness measurements had a 2 µm active layer (CC2233A), and those used for the mouse-specimen and therapeutic-energy measurements had a 2 µm thickness active layer (CC2233B); the array used for live monitoring had a 2 µm active layer (CC2233B).

### *Readout electronics*

As readout, an AMS-OSRAM AS5900 26-bit, 128-channel, low-noise current-to-digital converter was used, which implements a three-stage sigma-delta converter originally developed for X-ray applications. For the measurements presented here, the circuit was operated with an integration time of 1 ms and a clock frequency of 176.5 kHz. The digitised output was converted to physical units using the transfer relations provided in the device data sheet. One acquisition consists of a number of consecutive 1 ms frames, set by the chosen measurement length.

The GaN diode arrays were mounted on a printed circuit board (PCB) by adhesive bonding, and individual diodes were connected to the PCB readout traces via direct gold wire bonding. The readout electronics were positioned outside the beam path to minimise radiation exposure. Signals were routed from the diodes to the readout circuit via PCB traces on the sensor board and flex-connector cables to the readout module. The diodes were operated under near-zero-bias conditions. The bias voltage was externally applied to the AMS chip by using a regulated DC source of the model type KIKUSUI PMX18-5A.

### *Proton irradiation setup at CYRCé*

Experiments were performed at the “Plateforme de Radiobiologie Expérimentale de CYRCé” (PRECy) beamline of the “Cyclotron pour la recherche et l’enseignement” (CYRCé) facility in Strasbourg, which delivers protons with energies up to 25 MeV at beam currents up to the tens-of-nanoampere range. The beam was homogenised by inserting an aluminium degrader foil 344.5 cm upstream of the beam exit window. The beam current was monitored using a Faraday cup 47.4 cm upstream of the beam exit window. A calibration factor from the Faraday cup to a round aperture behind the beam exit window is utilised as specified by the CYRCé facility. Beam current and energy were adjustable remotely from the control room, enabling online adjustment during irradiation.

The detector was housed in an experimental chamber positioned at the beam exit. The chamber was maintained at a moderate vacuum to prevent the accumulation of beam-activated or chemically reactive air species, and was sealed against ambient light to avoid photocurrent contributions to the detector signal. Absorber phantoms could be inserted between the beam exit and the detector within the chamber. The detector was mounted on a motorised translation stage with sub-millimetre positioning precision, placed as close as practically possible to the phantom, with a nominal detector-to-phantom distance of below 10 mm. The detector position was calibrated relative to the beam axis using a cross-hair reference marker on the rear face of the PCB, aligned with the beam centre using an optical telescope. A lead block positioned behind the detector and PCB assembly provided beam stopping downstream of the active detector volume. The readout electronics were located outside the beamline and controlled remotely via a computer interface.

Further specifications of experimental setup and geometry are given for the individual measurement campaigns.

### *Signal quality and uniformity measurements*

Linearity measurements were performed using the homogenised beam (200 µm aluminium degrader foil, 36 mm aperture) at an incident proton energy of 25 MeV and a beam current of 2 nA at the Faraday cup. Although the beam is substantially homogenised over the aperture, residual spatial non-uniformity remains. To characterise the cross-channel response uniformity of the array, the detector was translated through the beam in steps equal to the array pitch, such that each spatial beam position was successively sampled by multiple diodes. This yields up to 128 independent signal comparisons per beam position, from which the channel-to-channel non-uniformity is quantified as the mean relative deviation of each channel's response from the global mean across all measured positions. The statistical within-channel signal fluctuation, assessed from repeated measurements at fixed positions, was found to be below 0.5% – smaller than the cross-channel spatial variation, confirming that the measured non-uniformity reflects genuine channel-to-channel differences rather than measurement noise. Slow beam current drifts were accounted for by subtracting a low-frequency background component from the array signal. The resulting non-uniformity distribution across the array is shown in Fig. 1h, with the inset displaying the corrected channel response.

### *Spatial-resolution and biological-imaging measurements*

The beam was homogenised using a 200 µm aluminium degrader, with a proton current around 2±0.1 nA measured at the Faraday cup. Imaging was performed with an 11 × 11 diode array (0.5 mm pixel pitch), translated across the field of view in steps smaller than the pixel pitch to spatially oversample the detector response and reconstruct an image at sub-pitch resolution.

Post-mortem mouse paws were collected from surplus animals euthanised in an authorized facility according to institutional and legal ethical guidelines. In accordance with French regulations, no formal project approval (APAFiS) was required for this post-mortem tissue use.

### *Radiochromic-film comparison*

Gafchromic EBT3 films (8 × 10 inch, Ashland LLC, Bridgewater, NJ, USA) were exposed at reduced beam current to achieve maximum optical contrast without film saturation. No absolute dose calibration was performed, as the films were used exclusively for qualitative spatial comparison with the detector array response; signal amplitude comparisons between film and detector were not intended. After exposure, films were digitised using a flatbed greyscale scanner (Epson Perfection V700 Photo). The greyscale response was normalised by subtracting the unexposed film background level, with the maximum acquired greyscale value defined as a normalised signal of 1 and the background level defined as 0. The inherent non-linearity of the radiochromic response between these limits is acceptable given the purely qualitative nature of the comparison. To reduce noise and match the spatial resolution of the detector array, the film response was spatially averaged to an effective pixel size of 250 µm.

Radiochromic film images and measurements were aligned by an optimisation process. The alignment transformation (comprising a lateral shift and rotation) was determined by minimising the root mean square difference between the overlapping image regions. To preserve the original spatial information, the transformation was applied to the full-resolution film image using nearest-neighbour interpolation only, with no higher-order resampling. The resulting aligned profiles share an identical spatial reference frame, enabling direct positional comparison. The detector signal was projected onto the interval [0, 1] by clipping any negative values (arising from dark current subtraction) to zero, and dividing by the maximum signal, yielding a normalised response directly comparable to the film data.

### *Energy-dependent response measurements*

The incident proton energy was varied by inserting aluminium absorber foils of different thicknesses into the beam path using a degrader wheel. The foils were introduced 8.95 mm downstream of the beam exit window within the experimental chamber. The detector was placed approximately 10 mm downstream with respect to the last metal aperture, which was 36.95 mm downstream of the beam exit window. How foil thickness corresponds to proton energy behind a 200 µm degrader or 400 µm degrader foil in the beamline is given in Table M.1. For each energy setting, the detector response was extracted as the median signal over the central region of the beam aperture, thereby avoiding the radially increasing beam inhomogeneity caused by stronger lateral scattering behind thicker absorber foils. The response was subsequently normalised to the incident proton fluence using the proton current density derived from the Faraday cup measurement.

Table M.1 gives the proton energies behind the energy modulation wheel as a function of aluminium foil thickness and degrader thickness in the beamline.

| Foil thickness (mm) | Energy, 200 µm degrader (MeV) | Energy, 400 µm degrader (MeV) |
|---|---|---|
| 0 | 23.69 | 22.75 |
| 0.147 | 22.98 | 22.02 |
| 0.29 | 22.31 | 21.32 |
| 0.431 | 21.57 | 20.55 |
| 0.568 | 20.91 | 19.87 |
| 0.701 | 20.19 | 19.11 |
| 0.842 | 19.44 | 18.33 |
| 0.99 | 18.67 | 17.52 |
| 1.133 | 17.81 | 16.62 |
| 1.278 | 16.98 | 15.74 |
| 1.424 | 16.06 | 14.76 |
| 1.564 | 15.16 | 13.8 |
| 1.714 | 14.15 | 12.7 |
| 1.862 | 13.07 | 11.52 |
| 2.007 | 12.01 | 10.34 |

| Foil thickness (mm) | Energy, 200 µm degrader (MeV) | Energy, 400 µm degrader (MeV) |
|---|---|---|
| 2.156 | 10.77 | 8.93 |
| 2.296 | 9.51 | 7.44 |
| 2.448 | 8 | 5.54 |
| 2.599 | 6.22 | 2.87 |
| 2.761 | 3.53 | 0 |

Two wedge absorber configurations were used for the energy characterisation measurements. The first wedge (14° opening angle, 40 × 40 × 10 mm³) was machined from polypropylene and placed downstream of a degrader wheel; the foil thicknesses and corresponding proton energies are specified in Table M.1. The usable beam aperture in this configuration was limited to approximately 26 mm in diameter by the degrader wheel, and measurements at the lowest proton energies were adversely affected by this aperture restriction and are therefore excluded from the analysis. A 200 µm aluminium degrader was installed in the beamline, and measurements were performed with a 2 × 64 pixel array (0.5 × 0.5 mm² pixel pitch), in which one pixel row is laterally offset by 0.25 mm relative to the other to enable sub-pitch spatial sampling.

The same detector array was used for measurements with the second wedge absorber (5° opening angle, 40 × 40 mm², thickness ranging from 2.14–5.64 mm), machined from polyetheretherketone (PEEK). This wedge was placed directly against the beam exit window with no air gap, minimising additional scattering and energy straggling between the wedge and the detector. The beamline was configured with a 40 µm aluminium degrader and a 50 µm exit window. The incident proton energy was varied at the cyclotron level across three settings: 23, 24 and 25 MeV.

### *Measurements at therapeutic energies*

Another measurement campaign was performed at the WPE at therapeutic proton energies. The shown values were measured at nominal beam energies (measured initial ranges R80) of 100 MeV (77.1 mm) to 220 MeV (306.1 mm) in 10 MeV steps and at the highest available energy of 226.7 MeV (322.29 mm) behind 2 cm of water-equivalent RW3 plates. The analysis of the data uses an estimate for the beam spot size to normalise to the current density.

### *GATE simulation setup*

Monte Carlo simulations were performed using the GATE software (OpenGate package v10.1.0)[62,63], which is based on Geant4 (v11.4.0)[64] as its simulation backend. All simulations were implemented and analysed in Python (v3.10.5). Particle physics interactions were modelled using the QGSP_BIC_EMZ physics list. Within the simulated detector structure, production cuts and maximum step sizes were both set to 0.1 µm to ensure accurate tracking of low-energy secondaries; outside the detector volume, production cuts were set to 10 cm to reduce computation time.

The geometry of the PRECy beamline at CYRCé was implemented to include the key beam-shaping elements: a scattering foil for beam homogenisation, positioned 344.5 cm upstream of the beam exit into air. No active collimation elements are present in the upstream beamline section; the primary beam was therefore modelled as a collimated beam of circular cross-section with 8 mm diameter, incident on the scattering foil. The beam exit window was simulated as 50 µm thick. The wheel was simulated with the above-described foil thicknesses. All aluminium pieces are simulated with 2.15 g $cm^{-3}$ density, which showed the best agreement with the given wheel energies (compare Table M.1).

The wedges were simulated with given dimensions out of polypropylene ($C_3H_6$, 0.94 g $cm^{-3}$) for the 14° wedge and out of PEEK ($C_{19}H_{12}O_3$, 1.31 g $cm^{-3}$) for the 5° wedge. Simulations involving wedge absorbers reproduced the wedge geometry according to the fabrication specifications, including the wedge angle, dimensions and material composition.

The detector was placed at the experimentally measured distance from the beam exit and modelled layer by layer, as described in the Detector fabrication section. To this end, each layer was modeled as a single volume with its correct thickness and material, while the simulation software was configured to separately track signals in individual detector pixels. The simulated detector structure comprised the complete layer stack of each diode up to the gold contact (excluding the wiring). Sapphire wafer and PCB board are included behind the detector.

No additional modelling of the conversion of deposited energy into electrical current through semiconductor processes or readout electronics was included, as the purpose of the simulation was to assess whether the deposited energy translates into the measured detector signal regardless of the intermediate processes. Accordingly, the quantities of interest were the dose and the energy deposited by protons arriving in the active i-GaN layer.

Proton fluence and energy distributions were extracted using OpenGate PhaseSpaceActors attached to key volumes in the simulation geometry. The proton energy spectrum entering the active detector layer was recorded at the entrance surface of the i-GaN volume. The total proton fluence entering the experimental setup was scored at two planes: directly upstream of the phantom, and at 47.4 cm from the beam exit into air, corresponding to the position of the Faraday cup in the experimental setup.

The dose deposited in the active detector layer was scored using a DoseActor attached to the i-GaN volume of the simulated diode structure. The actor was configured with a 200 × 200 pixel grid over a 50 × 50 mm² plane, oversampling the physical detector pixel pitch to provide sub-pixel spatial resolution. For comparison with experimental data, the simulated dose maps were subsequently rebinned to match the detector geometry.

### *Live monitoring and movie reconstruction*

Live beam monitoring was demonstrated using a fixed 11 × 11 pixel array with a square pixel pitch of 1 mm. The readout circuit was operated under the settings described above, with a continuous acquisition time of 1 s per data file repeated across multiple files. Measurements displayed on a millisecond timescale are therefore drawn from within a single data file, while those displayed on a seconds timescale span multiple consecutive files.

The video reconstruction was implemented in Python (v3.10.5), using ffmpeg (v7.1.1) for movie generation and compression. All movies are displayed at 60 frames per second; the number of measurement frames averaged per displayed frame is adjusted by the script to match the desired temporal resolution of the movie. Both the timing of individual measurements within a file and the inter-file acquisition gaps are accounted for to ensure accurate temporal reconstruction of the beam dynamics.

The measurements show the unhomogenised beam of the CYRCé facility, whose spatial profile is strongly dependent on the collimation settings of the beamline. Beam collimation and intensity were adjustable online during irradiation. The dynamic beam structure visible in the measurements reflects the operation of the x- and y-axis wire profilers installed within the PRECy beamline: as each wire is swept transversely through the beam, it attenuates the local beam intensity, generating a one-dimensional profile. The passage of the wire across the beam aperture is directly visualised in the detector response.

### *Long-term irradiation campaigns*

For the first long-term irradiation campaign, the detector was positioned as close as practicable to the beam exit window. The cyclotron was operated at an initial proton energy of 25 MeV, with the beam homogenised by a 20 µm degrader foil installed in the beamline. The beam current was set to 25 nA as measured at the Faraday cup; based on the conversion factor provided by the facility, 68.8% of this current reached the detector position. The beam current was monitored regularly throughout the campaign, and periodic beam interruptions were accounted for in the accumulated dose and fluence calculations. The signal of a single central diode was recorded as a function of irradiation time. Array linearity was characterised before and after the campaign. The cumulative dose of 2.4 MGy was estimated from the proton stopping power at the given energy, combined with the irradiation time, current density, and diode geometry; as this approach assumes complete energy deposition within the active layer, it represents a conservative estimation of the actual deposited dose.

The second campaign introduced a 5° PEEK wedge absorber (as described in the Energy-dependent response measurements section) placed directly at the beam exit window. The 2 × 64 pixel array (0.5 × 0.5 mm² pixel pitch, 0.25 mm inter-row offset) was mounted in a fixed position 10 mm in front of the wedge. Beam homogenisation was achieved with a 40 µm aluminium degrader foil, and the standard exit window was replaced with a 50 µm aluminium foil. The beam current was initially set to 8 nA at the Faraday cup for the first 40 min of irradiation, then increased to 25 nA; the facility-measured conversion factor from Faraday cup to phantom aperture was 88.74%. The total irradiation was carried out over two days, with a combined active irradiation time of approximately 15 hours; extended interruptions between sessions were accounted for in all cumulative dose and fluence calculations. For display purposes, the data acquired at 8 nA are shown rescaled to 25 nA equivalent, but all quantitative calculations use the original measured current values. Array linearity was characterised before and after the campaign.

Proton energy, fluence and cumulative dose per diode were derived from GATE Monte Carlo simulations as described in the GATE simulation setup section. NIEL values were calculated using the SR-NIEL online calculator[66]. For the Ga sublattice a displacement threshold of $E_d$=19 eV was assumed[45], for the N sublattice a value of $E_d$=17 eV[69].

## Data availability

The source data underlying the findings of this study are available in the Zenodo repository under DOI 10.5281/zenodo.22200263 (https://doi.org/10.5281/zenodo.22200263) or title "MATRIX GaN diode-array proton beam characterisation".

## Code availability

The used simulation platform GATE 10 is freely available on GitHub under https://github.com/OpenGATE/opengate.

## Acknowledgements

The authors acknowledge funding by the Deutsche Forschungsgemeinschaft (DFG, German Research Foundation) Project No. 505408069, funding by the Agence Nationale de la Recherche (Matrix, ANR-22-CE92-0047), NECTAR (ITMI/INSERM-2021), and DFH/UFA CDFA-05-06 Doctoral School. Matilde Siviero acknowledges funding from the French-German University/Saarbrücken (contract CDOC-06-2022). We gratefully acknowledge ams OSRAM for providing the AS5900 readout circuit used for data acquisition. The authors acknowledge the support of M. Pellicioli and J. Schuler during experiments at CYRCé.

## Additional Information

J.Y.D., M.H., C.B. and A.D.W. planned the project and applied for funding. E.F. grew the GaN structures, and M.S. fabricated the pixel arrays. S.H. developed the data-acquisition system spanning the detector readout, measurement and initial analysis, and carried out and analysed the initial tests. S.H. and N.B. performed the experiments at CYRCé, with support from J.Y.D., M.S. and M.H. N.B. analysed the data, and all authors interpreted and discussed the results. C.B. and L.P. supervised the experiments at the WPE. N.B. performed the simulations under the supervision of N.A. N.B. prepared the manuscript. All authors reviewed and approved the final manuscript.

## Competing interests

The authors declare no competing interests.